\documentclass{ws-ijmpb}

\begin{document}

\markboth{Hui Zhai}
{Spin-Orbit Coupled Quantum Gases}

%%%%%%%%%%%%%%%%%%%%% Publisher's Area please ignore %%%%%%%%%%%%%%%
%
%\catchline{}{}{}{}{}
%
%%%%%%%%%%%%%%%%%%%%%%%%%%%%%%%%%%%%%%%%%%%%%%%%%%%%%%%%%%%%%%%%%%%%

\title{Spin-Orbit Coupled Quantum Gases  }

\author{Hui Zhai}

\address{Institute for Advanced Study, Tsinghua University, Beijing, 100084, China  }

\maketitle

%\begin{history}
%\received{Day Month Year}
%\revised{Day Month Year}
%\accepted{(Day Month Year)}
%\comby{(xxxxxxxxxx)}
%\end{history}

\begin{abstract}

In this review we will discuss the experimental and theoretical progresses in studying spin-orbit coupled degenerate atomic gases during the last two years. We shall first review a series of pioneering experiments in generating synthetic gauge potentials and spin-orbit coupling in atomic gases by engineering atom-light interaction. Realization of spin-orbit coupled quantum gases opens a new avenue in cold atom physics, and also brings out a lot of new physical problems. In particular, the interplay between spin-orbit coupling and inter-atomic interaction leads to many intriguing phenomena. Here, by reviewing recent theoretical studies of both interacting bosons and fermions with isotropic Rashba spin-orbit coupling, the key message delivered here is that spin-orbit coupling can enhance the interaction effects, and make the interaction effects much more dramatic even in the weakly interacting regime.


\end{abstract}

\vspace{0.1in}

Many interesting phenomena in condensed matter physics occur when electrons are placed in an electric or magnetic field, or possess strong spin-orbit (SO) coupling. However, in the cold atom systems, neutral atoms neither possess gauge coupling to electromagnetic fields nor have SO coupling. Recently, by controlling atom-light interaction, one can generate an artificial external abelian or non-abelian gauge potential coupled to neutral atoms. The basic principle is based on the Berry phase effect \cite{Berry} and its non-abelian generalization \cite{Wilczek}. An important application of this scheme is creating an effective SO coupling in degenerate atomic gases. Since 2009, Spielman's group in NIST has successfully implemented this principle and generated both synthetic uniform gauge field \cite{NIST_uniform}, magnetic field \cite{NIST_magnetic}, electric field \cite{NIST_electric} and SO coupling \cite{NIST_spin_orbit}. We shall discuss the experimental progresses along this line in Sec. \ref{SO}.

The effects of SO coupling in electronic systems have been extensively discussed in condensed matter physics before and are also important topics nowadays. One of the most famous example is recently discovered topological insulators \cite{TI_1,TI_2,TI_3}. In this review, we try to convey the message that SO coupling in degenerate atomic gases will bring out new physics which have not been considered before, mainly due to the interplay between SO coupling and the unique properties of atomic gases. For bosonic atoms, SO coupled interacting bosons is a system never explored in physics before. For fermionic atoms, since a lot of intriguing physics have been revealed during the last ten years by utilizing Feshbach resonance technique to achieve interaction as strong as Fermi energy, the interplay between resonance physics and SO coupling is definitely a subject of great interests. 

A key point we want to emphasize in this review is that {\it a nearly isotropic SO coupling will dramatically enhance the effects of inter-particle interactions, so that the interaction effects are not weak even in the regime where the interaction strength itself is small}. This is because an isotropic Rashba SO coupling or a nearly isotropic SO coupling significantly changes the low-energy states of single particle Hamiltonian, as we shall discuss in Sec \ref{Single}. In Sec. \ref{Boson}, we discuss many-body system of bosons. Because the single particle ground state has large degeneracy, it is the inter-particle interaction that selects out a unique many-body ground state and determines its low-energy fluctuations. In Sec. \ref{Fermion}, we discuss many-body system of fermions. Because the low-energy density-of-state (DOS) is largely enhanced, the interaction effects become much more profound, in particular for weak attractions. A brief summary and future perspective are given in Sec. \ref{Summary}. 

\section{Synthetic Gauge Potentials and Spin-Orbit Coupling \label{SO}}

\begin{figure}[htbt]
\centerline{\psfig{file=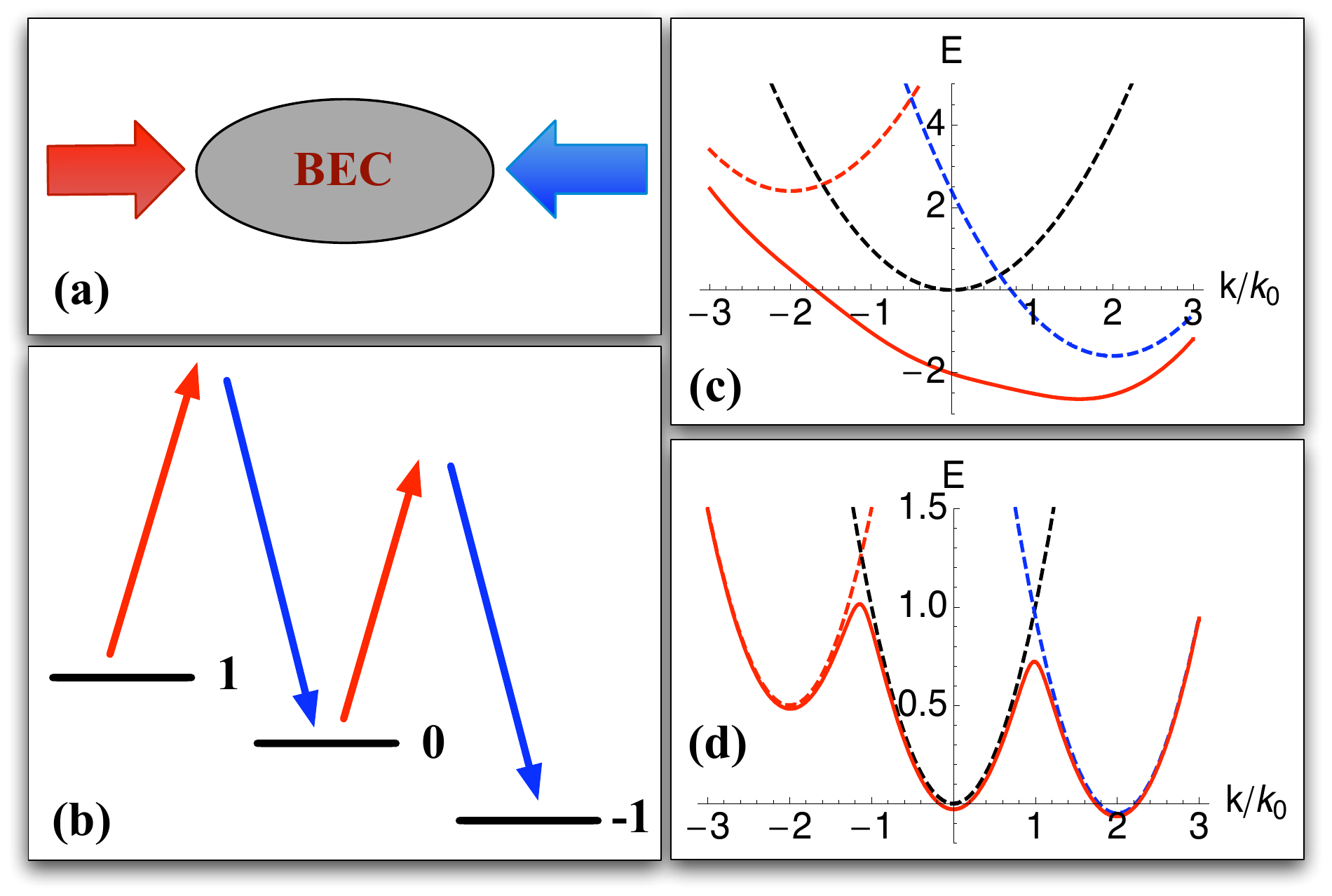,width=3.in}}
\vspace*{8pt}
\caption{(a) A schematic of NIST experiment, in which two counter propagating Raman beams are applied to $^{87}$Rb BEC. (b) A schematic of how three $F=1$ levels are coupled by Raman beams. (c) Dispersion in the regime of uniform vector potential. (d) Dispersion in the regime of non-abelian gauge field. \label{NIST}}
\end{figure}

In 2009, Spielman's group in NIST first realized a uniform vector potential in Bose-Einstein condensate (BEC) of $^{87}$Rb \cite{NIST_uniform}. In this experiment, two counter propagating Raman laser beams couple $|F,m_{\text{F}}\rangle=|1,-1\rangle$ level of $^{87}$Rb to $|1,0\rangle$ level, and couple $|1,0\rangle$ level to $|1,1\rangle$ level, as shown in Fig. \ref{NIST}(a) and (b), which can be described by the Hamiltonian
\begin{equation}
H=\left(\begin{array}{ccc} \frac{k^2_x}{2m}+\epsilon_1 & \frac{\Omega}{2}e^{i2k_0x} & 0 \\ \frac{\Omega}{2}e^{-i2k_0x} & \frac{k^2_x}{2m} & \frac{\Omega}{2}e^{i2k_0x} \\0 & \frac{\Omega}{2}e^{-i2k_0x} & \frac{k^2_x}{2m}-\epsilon_2 \end{array}\right)
\end{equation}
where $k_0=2\pi/\lambda$, $\lambda$ is the wave length of two lasers. $2k_0$ is therefore the momentum transfer during the two-photon processes. $\epsilon_1=\Delta_1+\delta\omega+\Delta_2$, and $\epsilon_2=\Delta_1+\delta\omega-\Delta_2$, where $\Delta_1$ denotes the linear Zeeman energy, $\delta\omega$ denotes the frequency difference of two Raman lasers, and $\Delta_2$ is the quadratic Zeeman energy. 

Applying a unitary transformation to wave function $\Phi=U\Psi$, where 
\begin{equation}
U=\left(\begin{array}{ccc}e^{-i2k_0 x} & 0 & 0 \\0 & 1 & 0 \\0 & 0 & e^{i2k_0x}\end{array}\right)
\end{equation} 
and the effective Hamiltonian becomes 
\begin{equation}
H_{\text{eff}}=UHU^\dag=\left(\begin{array}{ccc} \frac{(k_x+2k_0)^2}{2m}+\epsilon_1 & \frac{\Omega}{2} & 0 \\ \frac{\Omega}{2} & \frac{k^2_x}{2m} & \frac{\Omega}{2} \\0 & \frac{\Omega}{2} & \frac{(k_x-2k_0)^2}{2m}-\epsilon_2 \end{array}\right).
\end{equation}
When both $\epsilon_1$ and $\epsilon_2$ are large, the single particle energy dispersion of $H_{\text{eff}}$ is shown as Fig. \ref{NIST}(c), which displays a single energy minimum at finite $k_x$. In this regime the low energy physics is dominated by a single dressed state described by $\frac{1}{2m}(k_x-A_x)^2$, where $A_x$ is a constant. This leads to a uniform vector gauge field. 

In a follow up experiment, Spielman's group applied a Zeeman field gradient along $\hat{y}$ direction to this system  \cite{NIST_magnetic}. In this case, $A_x$ becomes a function of $y$ instead of a constant. It gives rise to a non-zero synthetic magnetic field $B_z=-\partial_y A_x\neq 0$. They observed a critical magnetic field above which many vortices are generated in the BEC \cite{NIST_magnetic}. In another experiment \cite{NIST_electric}, they made $A_x$ time dependent which gives rise to a non-zero electric filed $E_x=-\partial_t A_x\neq 0$. They observed collective oscillation of BEC after a pulse of electric field \cite{NIST_electric}.

By tuning the Zeeman energy and the laser frequency, one can also reach the regime where $\Delta_1+\delta\omega\approx\Delta_2$, and thus $\epsilon_2\approx 0$, while $\epsilon_1\approx 2\Delta_2$ is still large. In 2011, Spielman's group first reached this regime and showed that a SO coupling can be generated\cite{NIST_spin_orbit}. As shown in Fig. \ref{NIST}(d), in this regime the low-energy physics contains two energy minima which are dominated by $|1,-1\rangle$ and $|1,0\rangle$ states, respectively.  Therefore we can deduce the low-energy effective Hamiltonian by keeping both $|1,-1\rangle$  and $|1,0\rangle$, and rewrite the Hamiltonian as
\begin{equation}
H=\left(\begin{array}{cc} \frac{k^2_x}{2m}+\frac{h}{2} & \frac{\Omega}{2}e^{i2k_0x}   \\  \frac{\Omega}{2}e^{-i2k_0x}  & \frac{k^2_x}{2m}-\frac{h}{2} \end{array}\right)
\end{equation}
where $h=\epsilon_2$. Similarly, by applying a unitary transformation to the wave function with 
\begin{equation}
U=\left(\begin{array}{cc}e^{-ik_0 x} & 0 \\0 & e^{i k_0 x}\end{array}\right)
\end{equation}
one reaches an effective Hamiltonian that describes SO coupling
\begin{equation}
H_{\text{SO}}=UHU^\dag=\left(\begin{array}{cc} \frac{(k_x+k_0)^2}{2m}+\frac{h}{2} & \frac{\Omega}{2}  \\  \frac{\Omega}{2}  & \frac{(k_x-k_0)^2}{2m}-\frac{h}{2} \end{array}\right)=\frac{1}{2m}(k_x+k_0\sigma_z)^2+\frac{\Omega}{2}\sigma_x+\frac{h}{2}\sigma_z. 
\end{equation}
In fact, upon a pseudo-spin rotation $\sigma_x\rightarrow -\sigma_z$ and $\sigma_z\rightarrow \sigma_x$, the above Hamiltonian is equivalent to
\begin{equation}
H_{\text{SO}}=\frac{1}{2m}(k_x+k_0\sigma_x)^2-\frac{\Omega}{2}\sigma_z+\frac{h}{2}\sigma_x, \label{SO_NIST}
\end{equation}
where the first term can be viewed as an equal weight of Rashba ($k_x\sigma_x+k_y\sigma_y$) and Dresselhaus ($k_x\sigma_x-k_y\sigma_y$) SO coupling \footnote{In many literature, Rashba SO coupling denotes $k_x\sigma_y-k_y\sigma_x$, while Dresselhaus SO coupling denotes $k_x\sigma_y+k_y\sigma_x$. They are equivalent to the notations used in this paper by a pseudo-spin rotation $\sigma_x\rightarrow -\sigma_y$ and $\sigma_y\rightarrow \sigma_x$.}. The Hamiltonian of Eq. \ref{SO_NIST} can also be viewed as a Hamiltonian with synthetic non-abelian gauge field, since the vector potential $A_x=-k_0\sigma_z$ does not commute with the scale potential $\Phi=\frac{\Omega}{2}\sigma_x+\frac{h}{2}\sigma_z$.

Later on, Campbell {\it et al.} discussed how to generalize the NIST scheme to create SO coupling in both $\hat{x}$ and $\hat{y}$ direction, and nearly isotropic Rashba SO coupling \cite{Spielman_boson}. Xu and You recently introduce a dynamics generalization of the NIST scheme \cite{You}. Sau {\it et al.} discussed an explicit method to create Rashba SO coupling in fermionic $^{40}$K system at finite magnetic field, where a magnetic Feshbach resonance is available \cite{fermion}. Beside the NIST scheme, there are also other theoretical proposals that use $\Lambda$-type or tripod system to generate synthetic magnetic field \cite{Dark1,Dark2,Dark3,Dark4}, non-abelian gauge field with a monopole \cite{non-abelian} and SO coupling \cite{spin-orbit1,spin-orbit2,spin-orbit3,spin-orbit4}. However, for those proposals using dark states \cite{Dark1,Dark2,Dark3,Dark4,spin-orbit3,spin-orbit4}, collisional stability is a concern since there is always one eigen-state whose energy is below the dark state manifold, and multi-particle collision can lead to decay out of the dark state manifold. A recent review paper by Dalibard {\it et al.} has discussed different proposals in detail \cite{review}. 

\section{Single Particle Properties with Rashba Spin-Orbit Coupling \label{Single}} 

In the rest part of this review we will consider SO coupling in both $\hat{x}$ and $\hat{y}$ directions, whose Hamiltonian is given by 
\begin{equation}
H_0=\frac{1}{2m}\left[(k_x-\kappa_x\sigma_x)^2+(k_y-\kappa_y\sigma_y)^2+k^2_z\right]
\end{equation}
and in particular, we will consider the most symmetric Rashba case $\kappa_x=\kappa_y=\kappa>0$, where the physics is most interesting. In this case, the Hamiltonian can be rewritten as
\begin{equation}
H_0=\frac{1}{2m}\left({\bf k}_{\perp}^2-2\kappa{\bf k}_{\perp}\cdot{\bf \sigma}+\kappa^2+k^2_z\right)
\end{equation}
Obviously, spin is no longer a good quantum number. However ``helicity" is a good quantum number. ``Helicity" $\pm$ means that the spin direction is either parallel or anti-parallel to the in-plane momentum direction. For these two helicity branches, their dispersion are given by 
\begin{equation}
\epsilon_{{\bf k}\pm}=\frac{1}{2m}(k_{\perp}^2\mp 2\kappa k_{\perp}+\kappa^2+k^2_z)
\end{equation}  
where ${\bf k}_{\perp}=(k_x,k_y)$ and $k_{\perp}=\sqrt{k^2_x+k^2_y}$. This Hamiltonian displays a symmetry of simultaneously rotation of spin and momentum along $\hat{z}$ direction.

\begin{figure}[tbt]
\centerline{\psfig{file=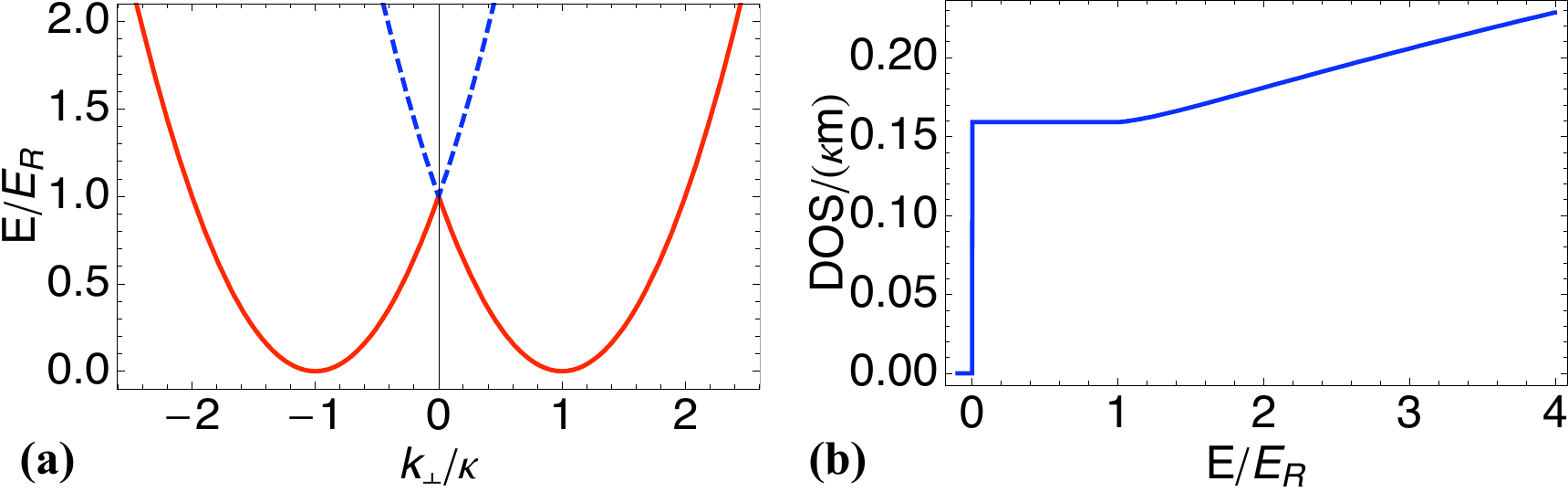,width=4.5in}}
\vspace*{8pt}
\caption{Energy dispersion with $k_z=0$ (a) and density-of-state (b) for Rashba spin-orbit coupled single particle Hamiltonian. In (a), ``helicity" is $``+"$ for red solid line and is $``-"$ for blue dashed line. $E_{\text{R}}=\kappa^2/(2m)$ is introduced as energy unit.    \label{single_particle}}
\end{figure}

As shown in Fig. \ref{single_particle}(a), helicity plus branch has lower energy. The single particle energy minimum has finite in-plane momentum $k_{\perp}=\kappa$, and all the single particle states with same $k_{\perp}=\kappa$ and $k_z=0$ but different azimuthal angle are degenerate ground states. The single particle DOS also has non-trival feature. In a three-dimensional system without SO coupling, DOS vanishes as $\sqrt{\epsilon}$ at low energies. However, in this system, as shown in Fig. \ref{single_particle}(b), the low-energy DOS becomes a constant when $\epsilon<E_{\text{R}}$, similar to a conventional two-dimensional system. 

Without SO coupling, the single particle Hamiltonian has a unique ground state at ${\bf k}=0$. At zero-temperature, bosons are all condensed at ${\bf k}=0$ state. If the interaction is weak, interaction effect is perturbative which only creates finite quantum depletion to the zero-momentum condensate. Also, because the vanishing DOS at low energies, two-body bound state appears only when the attractive interaction is above a threshold. Far below the threshold, when the attractive interaction is weak, the strength of fermion pairing and the fermion superfluid transition temperature are both exponentially small. That is to say, without SO coupling, not surprisingly, the interaction effect is weak in the regime where the interaction strength is small. 

With SO coupling, the effects of interaction are significantly enhanced even when its strength is still weak, because SO coupling significantly changes the single particle behaviors as discussed above. First, because the single particle ground state is not unique now, the ground state of a boson condensate is also not unique if there is no interactions. That is to say, it is the interaction that selects out a unique ground state among many possibilities. In this sense, the effect of interaction is non-perturbative even for very weak interactions. Secondly, because the DOS is a constant at low energies, a two-body bound state appears for any weak attractions. Therefore, for weak attractive interactions, both pairing gap and superfluid transition temperature are largely enhanced by SO coupling. In particular, the superfluid transition temperature can reach the same order as the Fermi temperature even for weak attractions. These points will be discussed in the next two sections in more detail.

On the other hand, we shall always keep in mind that in real experiment, one can never achieve a perfect isotropic Rashba SO coupling. Therefore, in our following discussions, we  always first obtain interesting results in the isotropic Rashba limit, and then discuss how robust the results are if there is a little anisotropy, namely, a slight difference between $\kappa_x$ and $\kappa_y$. 

\section{Spin-Orbit Coupled Bose Gases\label{Boson}}

\begin{figure}[tbt]
\centerline{\psfig{file=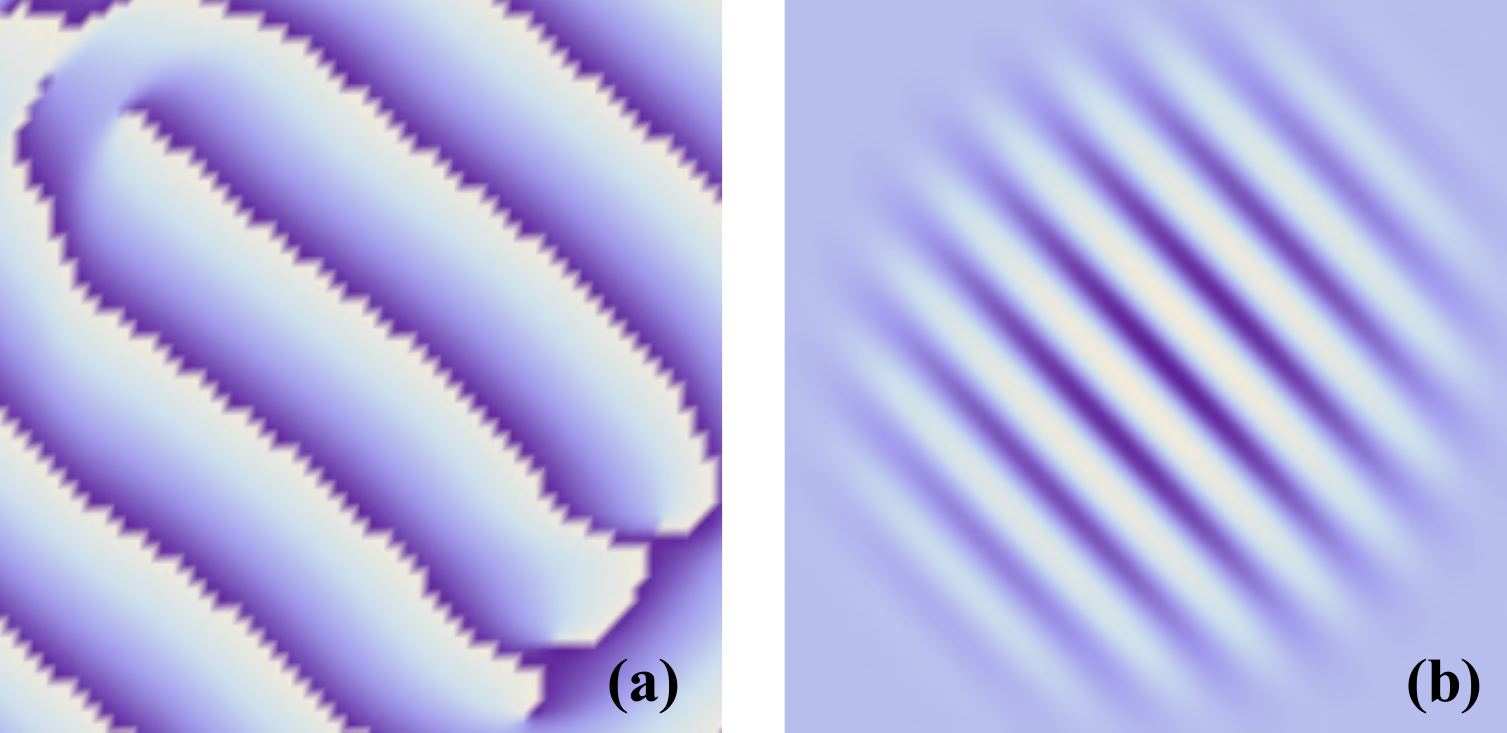,width=2.5in}}
\vspace*{8pt}
\caption{(a) Phase of condensate wave function in the ``plane wave phase"; (b) Spin density in the ``stripe superfluid" phase.  \label{mean-field}}
\end{figure}

We will first consider (pseudo-)spin-$1/2$ bosons. We shall first consider the mean-field ground state \footnote{There are also proposals of non-mean-field fragmented state, like N00N state, as ground state of SO coupled bosons \cite{Galitski}, however, the conventional wisdom is that such a state is very fragile when external perturbations are present. } and then discuss the effects of quantum and thermal fluctuations on top of mean-field saddle points. Let us first consider the most simplified form of interactions
\begin{equation}
\hat{H}_{\text{int}}=\int d^3{\bf r}\left(g n^2_1({\bf r})+g n^2_2({\bf r})+2g_{12}n_1({\bf r}) n_2({\bf r})\right) \label{int}
\end{equation}
By minimizing Gross-Pitaevskii energy functional, Wang {\it et al.} found that the mean-field ground state has two different phases depending on the sign of $g_{12}-g$ \cite{Wang}. When $g_{12}<g$, the system is in the ``plane wave phase", where all bosons are condensed into a single plane wave state, and the direction of plane wave is spontaneously chosen in the $xy$ plane. For instance, if the plane wave momentum is along $\hat{x}$ direction, the condensate wave function is given by
\begin{equation}
\psi=\sqrt{\frac{\rho}{2}}e^{i\kappa x}\left(\begin{array}{c}1 \\ 1\end{array}\right).
\end{equation}
The density of each component is uniform, but their phases modulate from zero to $2\pi$ periodically, as found from numerical solution of Gross-Pitaevskii equation and shown in Fig. \ref{mean-field}(a). This state spontaneously breaks time-reversal, rotational symmetry and the $U(1)$ symmetry of superfluid phase. When $g_{12}>g$, all bosons are condensed into a superposition of two plane wave states with opposite momentums, whose condensate wave function is given by
\begin{equation}
\psi=\frac{\sqrt{\rho}}{2}\left[e^{i\kappa x}\left(\begin{array}{c}1 \\ 1\end{array}\right)+e^{-i\kappa x}\left(\begin{array}{c}1 \\ -1\end{array}\right)\right]=\sqrt{\rho}\left(\begin{array}{c}\cos(\kappa x) \\i\sin(\kappa x)\end{array}\right)
\end{equation}
In this phase, the spin density $n_{1}-n_{2}=\rho\cos(2\kappa x)$ which has a periodic modulation in space, as shown in Fig. \ref{mean-field}(b), and therefore is named as ``stripe superfluid". The direction of the stripe is also spontaneously chosen in the $xy$ plane. Here, without loss of generality, we choose it along $\hat{x}$ direction. In this state, the high density regime of one component coincides with the low-density regime of the other component, so that the inter-component repulsive interaction is maximumly avoided. That is the reason why the spin stripe state is favored when $g_{12}$ is larger than $g$. In addition to the $U(1)$ superfluid phase, this state also breaks the rotational symmetry (but keeps the reflection symmetry), and the translational symmetry along the stripe direction. Hence, symmetry wise, it can also been called ``smectic superfluid". In contrast to the ``plane wave phase", this state does not break time-reversal symmetry. 

In principle, the condensate wave function can be any superposition of single particle ground states as
\begin{equation}
\psi=\sum_{\varphi}c_{\varphi}e^{i\kappa(\cos\varphi x+\sin\varphi y)}\left(\begin{array}{c}1 \\e^{i\varphi}\end{array}\right)
\end{equation}
One may wonder why only a single state or a superposition of a pair of states are favored by interactions. In fact, if there are more than a pair of single particle states in the superposition, the condensate wave function will exhibit interesting structure, such as various types of skyrmion lattices. And if all the degenerate states enter the condensate wave function with equal weight and specified relative phases, condensate will exhibit interesting structure of half vortices, as first proposed by Stanescu, Anderson, Galitski \cite{Galitski} and Wu, Mondragon-Shem \cite{CJ}. However, such a state is not energetically favorable in spin-$1/2$ case for a uniform (or nearly uniform) system. This is because that the interaction part can be rewritten as 
\begin{equation}
\hat{H}_{\text{int}}=\int d^3{\bf r}\left(\frac{g+g_{12}}{2}n^2({\bf r})+\frac{g-g_{12}}{2}S^2_z({\bf r})\right),
\end{equation}
where $n({\bf r})=n_{1}({\bf r})+n_{2}({\bf r})$ and $S_z=n_{1}({\bf r})-n_{2}({\bf r})$. The $S^2_z$-term can be satisfied by either choosing the ``plane wave phase" or the ``stripe phase", while the $n^2$-term with positive coefficient always favors a uniform density. One can easily show that, if there are more than a pair of states in the superposition, the condensate density will always have non-uniform modulation, which causes energy of $n^2$-term. Similar situation has also been found for spin-$1$ Hamiltonian \cite{spin-1,Wang}. However, there are several situations where skyrmion lattices or half vortices are found as ground state. When a strong harmonic confinement potential $V({\bf r})=m\omega r^2_{\perp}/2$ is applied to the system, condensate density is no longer uniform because of the trapping potential and the requirement from $n^2$-term becomes less restrictive. In addition, one notes in the limit of zero interaction, the ground state in a harmonic trap can be solved exactly and it is a half-vortex state \cite{Congjun,Hu,Santos}. Recently, three groups have found that if one largely increases the confinement potential so that $a=\sqrt{\hbar/m\omega}$ is comparable to $1/\kappa$, or reduces the interaction energy to be comparable to $\hbar\omega$, the ground state will evolve continuously to skyrmion lattice phases, and finally to half vortex phases \cite{Congjun,Hu,Santos}. Besides, Xu {\it et al.} and Kawakami {\it et al.} found in spin-$2$ case, because of an addition interaction term (which favors cyclic phase in absence of SO coupling \cite{spin-2,Ueda}), there exists a parameter regime where various types of skyrmion lattice phases are ground state even for a uniform system.

Back to the discussion of a nearly uniform spin-$1/2$ system\footnote{For typical experimental parameters, the BEC is in the nearly uniform regime rather than strong harmonic confinement regime.}, the ``plane wave phase" and the ``stripe phase" are in fact two very robust mean-field states. In real situation, the interactions between two pseudo-spin states have much more complicated form than the simplified form of Eq. (\ref{int}). Both Yip and Zhang {\it et al.} considered a specific type of complicated interaction form using a concrete realization of Rashba SO coupling, and they found the ground state is still either the ``plane wave phase" or the ``stripe phase" \cite{Yip,Zhang}. And also, the SO coupling is always not perfectly isotropic, say, $\kappa_x>\kappa_y$, then the Hamiltonian itself does not have rotational symmetry anymore. The single particle energy has two minima at $(\pm \kappa, 0,0)$ instead of a continuous circle of degenerate states. At mean-field level, the effect of anisotropic SO coupling is to pin the direction of plane wave or stripe into certain direction ($\hat{x}$ direction for this case). The NIST experimental situation discussed in Sec. \ref{SO} corresponds to the case $\kappa_y=0$ and also with a Zeeman field. As shown by Ho and Zhang \cite{Ho}, and also in the experimental paper \cite{NIST_spin_orbit}, the phase diagram of this system also only contains such two phases.  

To go beyond the mean-field description, three different approaches have been tried so far. The first is the effective field theory approach \cite{Jian}, which treats Gaussian fluctuations of low-energy modes. The second is Bogoliubov approach \cite{Qi,Biao,DaSarma}, which more focuses on the gapless phonon excitations. And the third is the renormalization approach \cite{Sarang,Gordom,Gordom2}, which discusses how scattering vertices are renormalized by high order processes. These different approaches address beyond-mean-field effects from different perspectives, and the results are consistent with each other where they overlap. 

Taking effective field theory approach as an example, for the ``stripe" phase,  the superfluid phase $\theta$ and the relative phase $u$ between two momentum components are two low-lying modes:
\begin{equation}
\varphi_{\text{ST}}=\frac{\sqrt{\rho}}{2}e^{i\theta}\left[e^{i(\kappa x+u)}\left(\begin{array}{c}1 \\1\end{array}\right)+e^{-i(\kappa x+u)}\left(\begin{array}{c}1 \\-1\end{array}\right)\right].
\end{equation}
In fact, the relative phase $u$ describes the phonon mode of stripe order. The dynamics of $\theta$ and $u$ fields are governed by an effective energy function \cite{Jian}
\begin{equation}
\mathcal{H}_{\text {eff}}^{{\text ST}}=\frac{\rho}{2m}\left[\left(\partial_x\theta\right)^2+\frac{\left(\partial_y\theta\right)^2}{\alpha^2}+
 \left(\partial_x u\right)^2+\frac{\left(\partial_y^2 u \right)^2}{4\kappa^2}
\right], \label{SWHeff}
\end{equation}
where $\alpha>1$ is a constant. For the ``plane wave" phase, the superfuid phase is the only low energy mode
\begin{equation}
\varphi_{\text{PW}}=\sqrt{\rho}e^{i(\kappa x+\theta)}\left(\begin{array}{c}1 \\1\end{array}\right)
\end{equation}
and its effective energy is derived as\cite{Jian}
\begin{equation}
\mathcal{H}_{\text {eff}}^{^{\text PW}}=\frac{\rho}{2m}\left[\left(\partial_x\theta\right)^2+\frac{1}{4\kappa^2}(\partial_y^2\theta)^2\right]
.\label{PWHeff}
\end{equation}
One notes that in the ``stripe" phase, the quadratic term $(\partial_y u)^2$ is absent in Eq. (\ref{SWHeff}), and in the ``plane wave" phase, the quadratic term $(\partial_y \theta)^2$ is absent in Eq. (\ref{PWHeff}). This is in fact a manifestation of rotational symmetry in this system. Similar effective theory has also been found for ``FFLO" state in fermion superfluid \cite{Leo_Ashvin,Leo}. Such an energy function is also the classical energy of smectic liquid crystal. 

In two dimensions, the finite temperature phase transition is driven by proliferation of topological defects. For usual $XY$ model, both the energy of topological vortex and its entropy logarithmically depend on system size. Thus, only above a critical temperature, entropy wins energy and the topological defects proliferate which drives system into normal phase. However, in the ``stripe" phase, because the absence of $(\partial_y u)^2$ term, the energy for a topological defect of $u$, i.e. a dislocation in the stripe, no longer logarithmically depends on system size. Hence it will lose to entropy at any finite temperature, and the proliferation of these dislocations will melt the stripe order. This restores translational symmetry and drives the system into a boson paired superfluid phase \cite{Jian}. Such a boson pairing state can also be predicted by looking at pairing instability of normal state with renormalized interactions \cite{Sarang}. 
%Moreover, it also indicates that the stripe correlation are both algebraically decayed at zero-temperature, namely, there is always no true long range order but has quasi-long-range order at zero temperature. The absence of true long range order can also deduced from the Bogoliubov approach as found recently by Zhou and Cui \cite{Qi}. 
By considering the renormalization effects of scattering vertex from high energy states,  Gopalakrishnan {\it et al.} \cite{Sarang} and Ozawa, Baym \cite{Gordom,Gordom2} found that the scattering amplitudes between two states with opposite momentum become smaller and even vanishing, which makes the ``stripe phase" more favorable at the low-density limit, and meanwhile leads to more significant the fluctuation of stripe order \cite{Sarang}. 
For same reason, in the ``plane wave" phase, the superfluid phase $\theta$ will immediately disorder at finite temperature and the system becomes normal.
% and also the superflud correlation is algebraically decayed at zero temperature \cite{Qi}.

In addition, the effective theory Eq.(\ref{SWHeff}) and Eq. (\ref{PWHeff}) also imply that there is a Goldstone mode which has linear dispersion along $\hat{x}$ direction (direction of stripe or plane wave momentum), and quadratic dispersion along $\hat{y}$ direction (direction perpendicular to the direction of stripe or plane wave momentum). Same results have been reached by Bogoliubov calculation \cite{Qi,Biao,DaSarma}. Similar behaviors of Goldstone modes also exist in ``FFLO" phase of fermion superfluid \cite{Leo_Ashvin,Leo_choi,Leo}.

When the SO coupling is anisotropic ($\kappa_x\neq \kappa_y$), there is no rotational symmetry in the Hamiltonian. Therefore, one will have $(\partial_y u)^2$ in the stripe phase, and $(\partial_y \theta)^2$ in the plane wave phase. 
%True long range order can now exist at zero-temperature \cite{Qi}. 
However, the coefficients of those terms are propositional to $1-(\kappa_y/\kappa_x)^2$. In the regime $\kappa_y/\kappa_x$ is very close to unity, the energy of a dislocation is much smaller than the energy of a vortex or a half vortex. Hence, the system will undergo two Kosterlitz-Thouless phase transitions. At a lower critical temperature, dislocations proliferate and the system becomes paired superfluid. Then, at a higher critical temperature, vortices proliferate and the system becomes normal. A completed phase diagram in term of interaction parameter, SO coupling anisotropy and temperature is given by Jian and Zhai \cite{Jian}.

For SO coupled bosons, many questions remain open. Recently several works begin to address the questions about the superfluid critical velocity \cite{Biao},  vortices in presence of external rotation \cite{Han,Congjun_2,Galitski_2}, the effects of dipolar interactions\cite{SuYi}, the collective modes \cite{Duine}, superfluid to Mott insulator transition in a lattice \cite{Lewenstein}, the interplay between magnetic field and SO coupling \cite{Burrello}, and the dynamical effects nearby Dirac point due to SO coupling \cite{Santos,Ohberg,Zhu}.  The research along this direction will definitely reveal more interesting physics and stimulate more interesting experiments. 

\section{Spin-Orbit Coupled Fermi Gases across a Feshbach Resonance \label{Fermion}}

Even for a non-interacting system, the thermodynamic behavior of a Fermi gas is dramatically changed by a strong SO coupling because of the change of the low-energy DOS \cite{Yu}. In this section, we focus on Fermi gas with attractive interaction, and in particular, across a Feshbach resonance. The interaction part is modeled as
\begin{equation}
H_{\text{int}}=g\int d^3{\bf r}\psi^\dag_{\uparrow}({\bf r})\psi^\dag_{\downarrow}({\bf r})\psi_{\downarrow}({\bf r})\psi_{\uparrow}({\bf r})
\end{equation} 
where
\begin{equation}
\frac{1}{g}=\frac{m}{4\pi\hbar^2 a_{\text{s}}}-\sum\limits_{{\bf k}}\frac{1}{\hbar^2{\bf k}^2/m}. \label{as}
\end{equation}
Here we relate the bare interaction $g$ to $s$-wave scattering length via Eq. (\ref{as}), which is the same as the scheme widely used in free space without SO coupling. This is based on the assumption that the interaction Hamiltonian is not changed by SO coupling, and the $a_{\text{s}}$ here should be understood as scattering length in free space. \footnote{This is equivalent to assume that the $s$-wave pseudo-potential is still a valid approximation for a short-range realistic potential in presence of SO coupling. In fact, the validity of such an approximation is not quite obvious, and it has only been examined recently by Cui \cite{Cui}.}

Using this interaction Hamiltonian,  Vyasanakere and Shenoy first studied the two-body problem across a Feshbach resonance \cite{Vijay-2body}. Because the low-energy DOS is now a constant, an arbitrary weak attractive interaction will give rise to a bound state. Similar situations are two-body problem in two-dimension and Cooper problem in three-dimension in absence of SO coupling, where DOS are also constants. The binding energy can be easily calculated by looking at poles of $T$-matrix \cite{Vijay-2body,Hu_fermion} or by reducing the two-body Schr\"odinger equation to a self-consistent equation \cite{Zhai}. The two-body properties at the BCS side and at resonance regime are significantly changed by SO coupling. At weakly interacting BCS side with small negative $\kappa a_{\text{s}}$, the binding energy behaves as
\begin{equation}
E_{\text{b}}\approx-\frac{\hbar^2\kappa^2}{2m}\frac{4}{e^2}e^{-\frac{2}{\kappa|a_{\text{s}}|}}.
\end{equation} 
where a large binding energy can always been reached by increasing the strength of SO coupling. 
At resonance with $a_{\text{s}}=\infty$, $1/\kappa$ is the only length scale in the two-body problem and therefore one has a universal result
\begin{equation}
E_{\text{b}}=-0.88\frac{\hbar^2\kappa^2}{2m}. 
\end{equation} 
While for the BEC side with small positive $\kappa a_{\text{s}}$, at leading order $E_{\text{b}}$ is still given by $\hbar^2/(2ma^2_{\text{s}})$ and is not affected by SO coupling. 
Moreover, because of SO coupling, the two-body wave function has both singlet and triplet components. For a two-body bound state with zero center-of-mass momentum, the wave function behaves as \cite{Vijay-2body}
\begin{equation}
\psi=\psi_{\text{s}}({\bf r})|\uparrow\downarrow-\downarrow\uparrow\rangle+\psi^*_{\text{a}}({\bf r})|\uparrow\uparrow\rangle+\psi_{\text{a}}({\bf r})|\downarrow\downarrow\rangle
\end{equation}
where $\psi_{\text{s}}({\bf r})$ and $\psi_{\text{a}}({\bf r})$ are symmetric and anti-symmetric functions, respectively. Furthermore, by looking at binding energy with finite center-of-mass momentum, one can determine the effective mass of molecules (two-body bound state). Hu {\it et al.} \cite{Hu_fermion} and Yu and Zhai \cite{Zhai} found that at the BCS limit, the effective mass of molecule finally saturates to $4m$, and at resonance, the effective mass is a universal number of $2.40m$. In the BEC limit, the effective mass saturates to $2m$ as conventional case.

\begin{figure}[tbt]
\centerline{\psfig{file=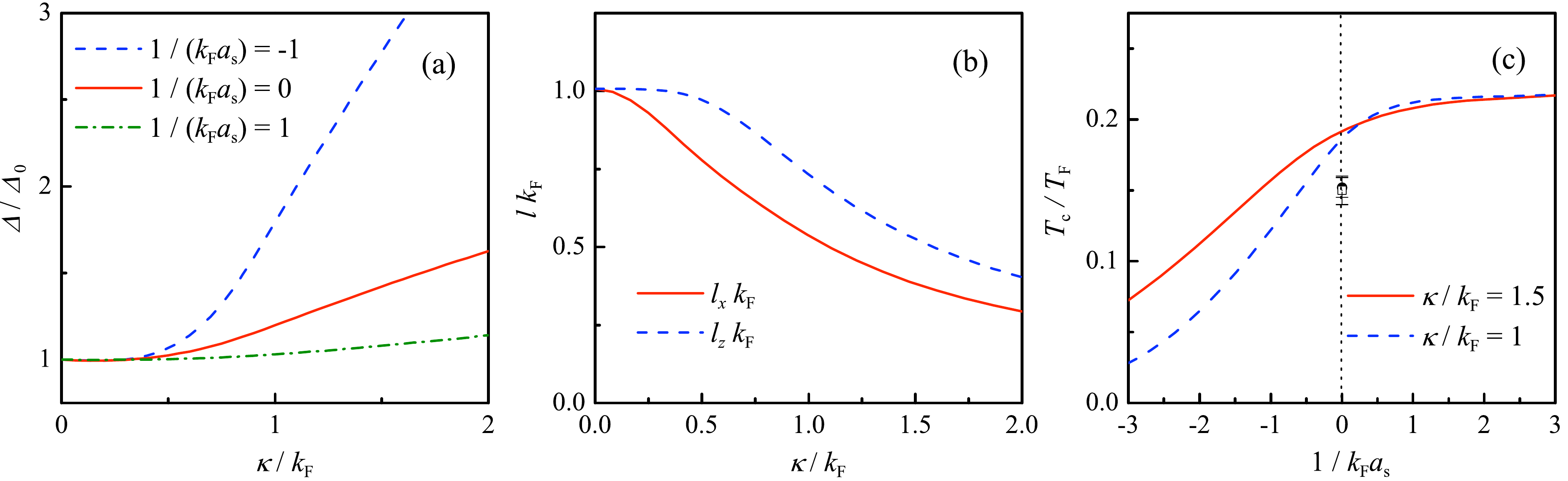,width=5.0in}}
\vspace*{8pt}
\caption{(a) Gap size $\Delta$ as a function of $\kappa/k_{\text{F}}$ for different values of $1/(k_{\text{F}}a_{\text{s}})$. (b) Size of Cooper pair in $xy$ plane $l_{\perp}$ and along $\hat{z}$ direction $l_z$ as a function of $\kappa/k_{\text{F}}$ at resonance with $a_{\text{s}}=\infty$; (c) Superfluid transition temperature $T_{\text{c}}/T_{\text{F}}$ as a function of $1/(k_{\text{F}}a_{\text{s}})$ for different values of $\kappa/k_{\text{F}}$. Reprinted from arXiv: 1105.2250 (Phys. Rev. Lett. to be published).  \label{fermion}}
\end{figure}

Generalizing conventional BEC-BCS crossover mean-field theory to the case with SO coupling and equal population $\langle n_{\uparrow}\rangle=\langle n_{\downarrow}\rangle$, one can show that the system remains gapped for all $k_{\text{F}}a_{\text{s}}$, although there are triplet $p$-wave components \cite{Vijay_mf,Hu_fermion,Zhai,Carlos_1}, and the pair wave function obtained from the mean-field theory finally approaches the wave function of two-body bound state (i.e. molecule wave function)\cite{Vijay_mf,Zhai}. Hence, it is still a crossover as $1/(k_{\text{F}}a_{\text{s}})$ changes from negative to positive. However, on the other hand, the change of low-energy DOS and the presence of two-body bound state at the BCS side and resonance regime will significantly change the properties of crossover  \cite{Vijay_mf,Hu_fermion,Zhai,Vijay3}. For example, as shown in Fig. \ref{fermion}(a), the pairing gap at the BCS side is dramatically enhanced when $\kappa/k_{\text{F}}$ is comparable or larger than unity \footnote{ In cold atom system, SO coupling is generated by atom-light coupling, and therefore $\kappa$ is on the order of the inverse of the laser wave length. And since the laser wave length and the inter-particle distance are comparable (between $\sim0.1\mu m$ and $\sim 1\mu m$) in atomic gases, the strength of SO coupling in cold atom systems can naturally reach the regime $\kappa/k_{\text{F}}\sim 1$.}. For such a strong SO coupling, the DOS at Fermi energy becomes a constant, and is much larger than the DOS in absence of SO coupling. That is the reason why the pairing effects become much dramatic even for same interaction strength. For another example,  in absence of SO coupling, Fermi energy is the only energy scale at resonance, and therefore the size of Cooper pair is a universal constant times $1/k_{\text{F}}$. SO coupling introduces another scale at resonance, which is $1/\kappa$. For large $\kappa$, as the pairing gap approaches two-body binding energy, the size of Cooper pairs also approaches $1/\kappa$. Fig. \ref{fermion}(b) shows that the behavior of $l$ undergoes a crossover from $1/k_{\text{F}}$ to $1/\kappa$ as $\kappa/k_{\text{F}}$ increases. This plot also shows that the size of Cooper pair in the $xy$ plane is different from the size along $\hat{z}$ direction, namely, the Cooper pairs are anisotropic.  \cite{Hu_fermion,Zhai}. In addition, one can also show that the superfluid transition temperature at the BCS side can also be enhanced a lot by SO coupling. For large enough SO coupling, it eventually approaches the BEC temperature of molecules with mass $4m$ \cite{Zhai,Vijay3}, which is a sizable fraction of Fermi temperature. The critical temperature across resonance is first estimated by Yu and Zhai \cite{Zhai} as shown in Fig. \ref{fermion}(c).

If SO coupling is slightly anisotropic, DOS at very low-energy will finally vanish. However, there is still a large energy window $\lesssim \hbar^2\kappa^2/(2m)$ where the DOS is greatly enhanced by SO coupling. Hence, pairing gap will still be enhanced as long as the density of fermions is not extremely low. Moreover, although it is no longer true that arbitrary small attraction can cause a bound state, the critical value for the appearance of a bound state will move from unitary point $a_{\text{s}}=\infty$ to the BCS side with negative $a_{\text{s}}$ \cite{Vijay-2body}. Once the bound state is present, it will influence the universal behavior of pair size at resonance and superfluid critical temperature as discussed above.  

For the imbalanced case, the phase diagram becomes more richer. Several groups have studied the phase diagram in presence of a Zeeman field \cite{Chuanwei,Iskin_1,YiWei,Carlos_2,Liao} and in various other circumstance \cite{Duan,Peter,Iskin_2,Iskin_3,Chuanwei_2,Dell,He,Vijay4,Yi,Duanfermion,Zhang} \footnote{ Iskin and Subasi studied SO coupled Fermi gas with mass imbalance \cite{Iskin_2}. They use mixture of different species as motivation of this study. We caution that the current way of generating SO coupling is based on light coupling of different internal state of atoms, which can not generate SO coupling if different internal states are different atomic species}. They have shown that, instead of a crossover, there are phase transitions between topological and non-topological phases \cite{Chuanwei,Iskin_1,YiWei,Carlos_2}. They have also discussed how SO coupling influences the competition between a uniform superfluid and a phase separation \cite{Iskin_1,YiWei}. %Moreover, Majorana fermion may be found in certain cases \cite{Duan,Peter}. 
%Recently, several groups have also discussed two-dimensional Fermi gas \cite{} and trapped Fermi gases \cite{} with SO coupling.

\section{Summary and Future Developments \label{Summary}}

SO coupled quantum gases with interactions are new systems in cold atom physics. Moreover, SO coupled bosonic system has never been thought in physics before. Currently our understanding of this system is still very limited, and many questions remain open. However, even from our limited experience with this new system, one can already get a feeling that this system has many unusual behaviors. This gives a lot of opportunities for theorists and experimentalists.

\section*{Acknowledgements}
I would like to thank Chao-Ming Jian and Zeng-Qing Yu for collaboration on this subject, and thank Xiaoling Cui for helpful discussions. This work is supported by Tsinghua University Initiative Scientific Research Program, NSFC under Grant No. 11004118 and No. 11174176, NKBRSFC under Grant No. 2011CB921500.

\end{document}